\documentclass[runningheads]{llncs}

\usepackage{graphicx}
\graphicspath{ {./images/} }
\usepackage{gensymb}
\usepackage{subcaption}
\begin{document}

\title{Art and Science Interaction Lab}

\subtitle{A highly flexible and modular interaction science research facility}
%
%
\author{Niels Van Kets\inst{1}\textsuperscript{\textsection}\orcidID{0001-5495-2240} \and
Bart Moens\inst{2}\textsuperscript{\textsection}\orcidID{0002-1281-9406} \and
Klaas Bombeke\inst{3}\textsuperscript{\textsection}\orcidID{0003-2056-1246} \and
Wouter Durnez\inst{3}\textsuperscript{\textsection}\orcidID{0001-8045-8801} \and
Pieter-Jan Maes\inst{2}\orcidID{0002-9237-3298} \and
Glenn Van Wallendael\inst{1}\textsuperscript{\textsection}\orcidID{0001-9530-3466} \and
Lieven De Marez\inst{3}\orcidID{0001-7716-4079} \and
Marc Leman\inst{2}\orcidID{0002-9780-2194} \and
Peter Lambert\inst{1}\textsuperscript{\textsection}\orcidID{0001-5313-4158}
}

\authorrunning{N. Van Kets, et al.}
%
\institute{imec-IDLab-UGent, Department of Electronics and Information Systems, Ghent University, Technologiepark-Zwijnaarde 126, 9052 Zwijnaarde, Belgium \and IPEM-UGent, Department of Art History, Musicology and Theatre Studies, Ghent University, Miriam Makebaplein 1, 9000 Gent, Belgium \and imec-mict-UGent, Department of Communication Sciences, Ghent University, Miriam Makebaplein 1, 9000 Gent, Belgium}
%


\maketitle              


\begin{abstract}

The Art and Science Interaction Lab (“ASIL”) is a unique, highly flexible and modular “interaction science” research facility to effectively bring, analyse and test experiences and interactions in mixed virtual/augmented contexts as well as to conduct research on next-gen immersive technologies. It brings together the expertise and creativity of engineers, performers, designers and scientists creating solutions and experiences shaping the lives of people. The lab is equipped with state-of-the-art visual, auditory and user-tracking equipment, fully synchronized and connected to a central backend. This synchronization allows for highly accurate multi-sensor measurements and analysis.

\end{abstract}



\section{Introduction}

The Art and Science Interaction Lab (ASIL) team supports innovation in different key domains. Within these domains, the team focuses on interaction research in virtualized environments, unraveling complex user interactions and experiences in order to design and create novel applications and interfaces. The application domains span from smart home appliances, health, safety, smart public places to more artistic and creative applications. Furthermore, the lab infrastructure is used for fundamental research on virtual reality technologies (e.g. auralization, virtual acoustics, 6 degrees of freedom VR, multi-person VR…) and is dark fiber connected to three concert halls in Ghent. \\ 

The team is an interdisciplinary consortium combining the expertise of three Ghent University research groups (IDLab, IPEM and mict) and has been co-funded under the medium-scale research infrastructure program governed by the Research Foundation Flanders (FWO). The combination of humanities, engineering, psychology and social sciences makes the ASIL a one of its kind research facility. Moreover, the research tracks in the ASIL target both industry and academia to deliver a unique interdisciplinary approach in measuring, analyzing and creating our next-generation appliances, interfaces and experiences. 


\section{Technical infrastructure}

\begin{figure}[!htbp]
    \centering
    \caption{De Krook}
    \includegraphics[width=\linewidth]{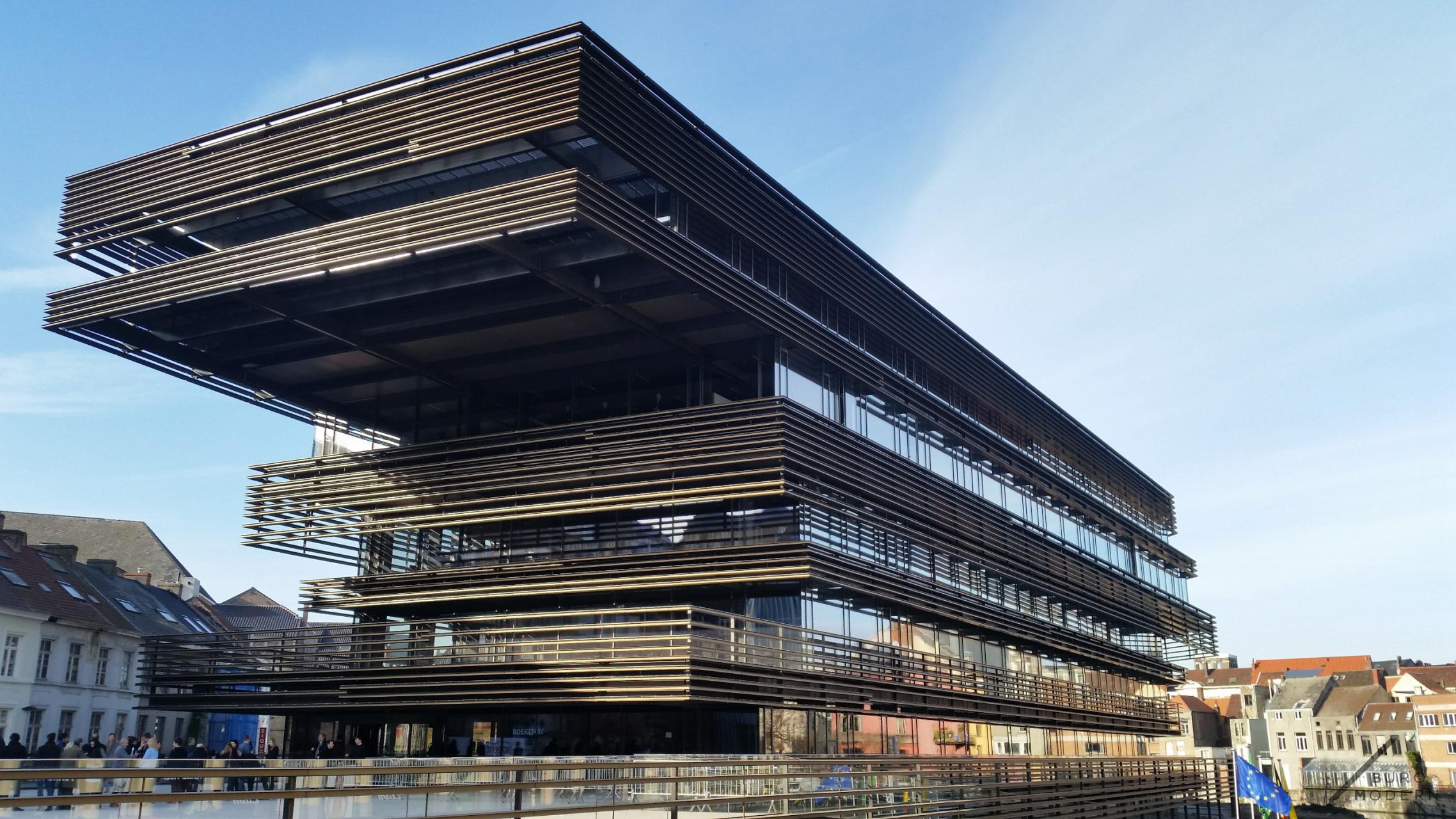}
    \label{fig:krook}
\end{figure}

The Art and Science Interaction lab is located in "De Krook" building (Fig. \ref{fig:krook}) in the city center of Ghent. Next to the city of Ghent library, "De Krook" also houses the three Ghent University research groups involved in the Art and Science Interaction Lab (IDLab, IPEM and mict), as well as imec, an R\&D hub for nano- and digital technologies. Both IDLab and mict are affiliated imec research groups. \\

The Lab is located in the sub-level floors of the building and has a volume of 10m x 10m and spans a height of two full floors ($\approx$ 6m). \\

In view of the audiovisual applications envisioned in this room, the walls and ceiling all have been acoustically treated, delivering a RT60 Reverberation Time of 0.5s. The RT60 Reverberation Time describes how long it takes for sound to decay by 60dB in a room with a diffuse soundfield. By insulating the room, a reduction of approximately 2s was met. \\

\begin{figure}[!htbp]
    \centering
        \subfloat[\centering Motorized Trusses]{{\includegraphics[width=5cm]{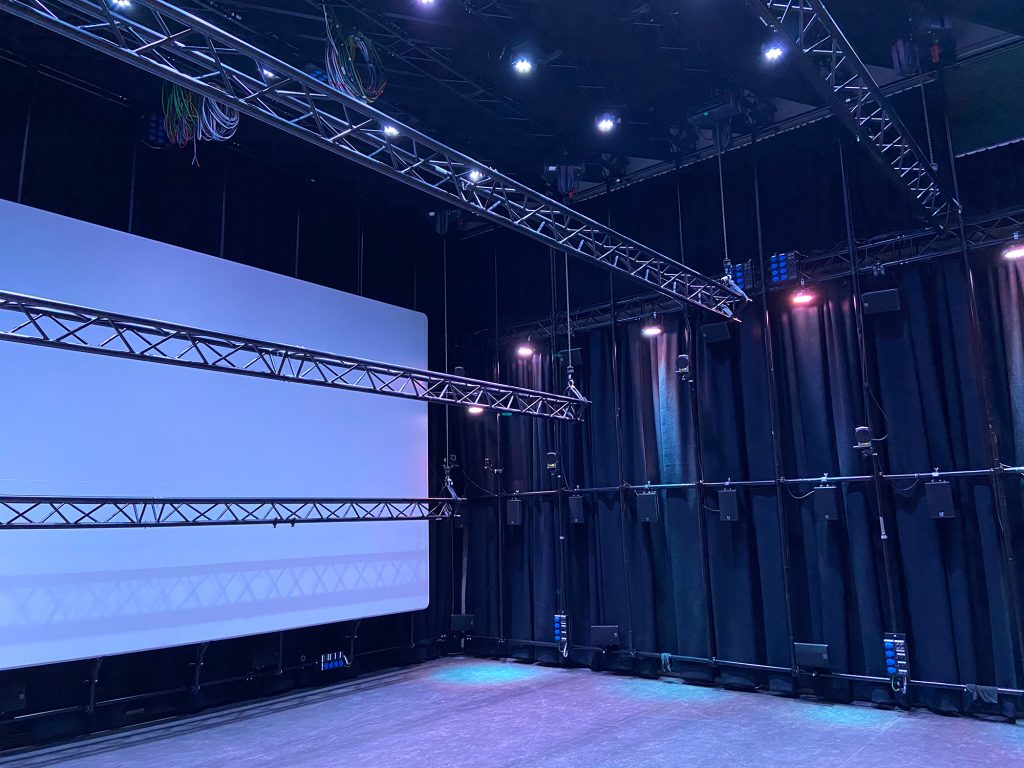} }}%
        \qquad
        \subfloat[\centering Overview]{{\includegraphics[width=5cm]{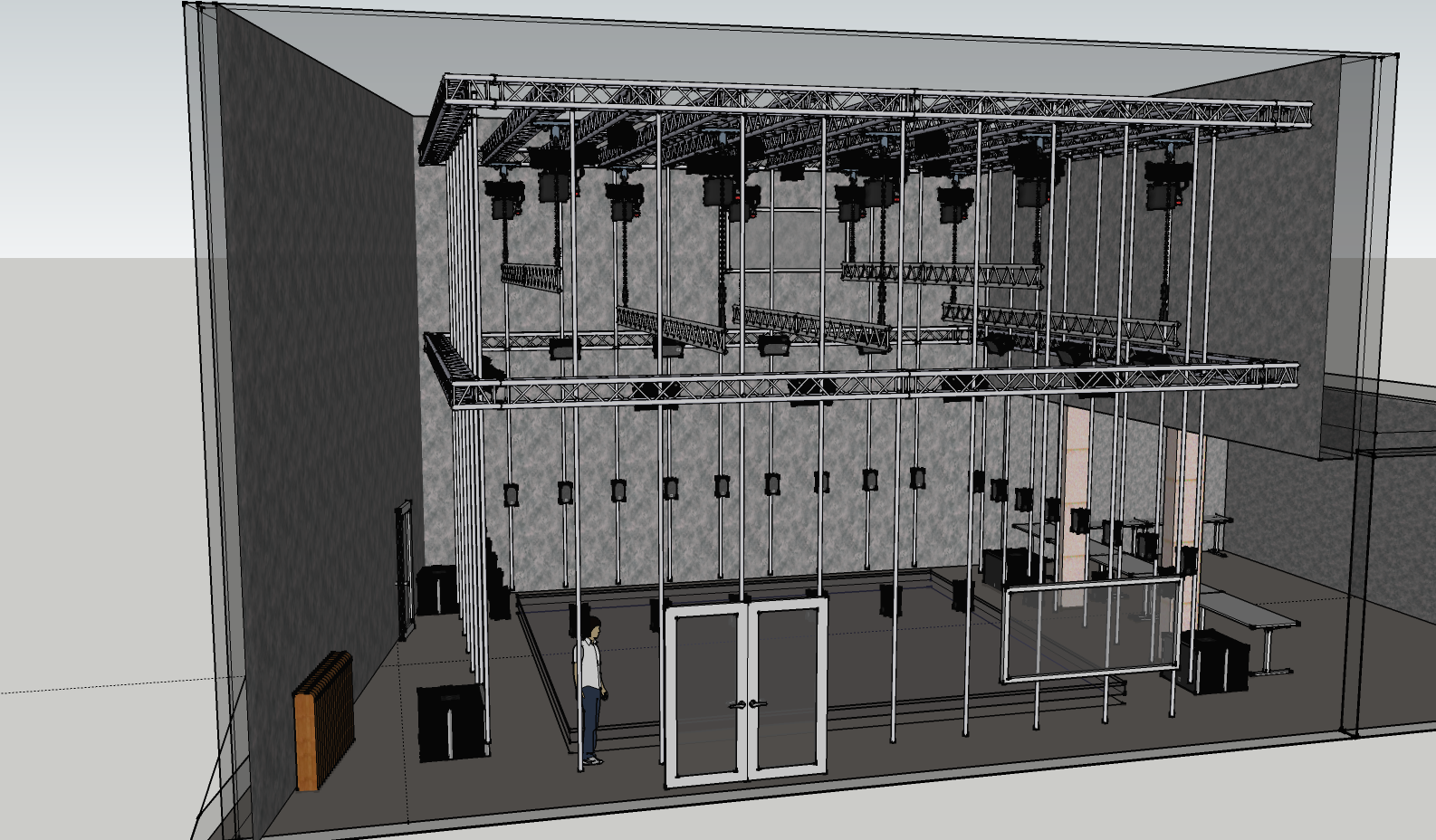} }}%
        \caption{Trussing system}%
        \label{fig:trussing}%
\end{figure}

Furthermore, in order to deliver a highly flexible and modular research infrastructure, the lab has been equipped with a state of the art trussing system (Fig. \ref{fig:trussing}, including 5 motorized trusses of 7m. This trussing system allows for a myriad of experimental setups. 

\begin{figure}[!htbp]
    \centering
        \subfloat[\centering Patch Box]{{\includegraphics[height=4cm]{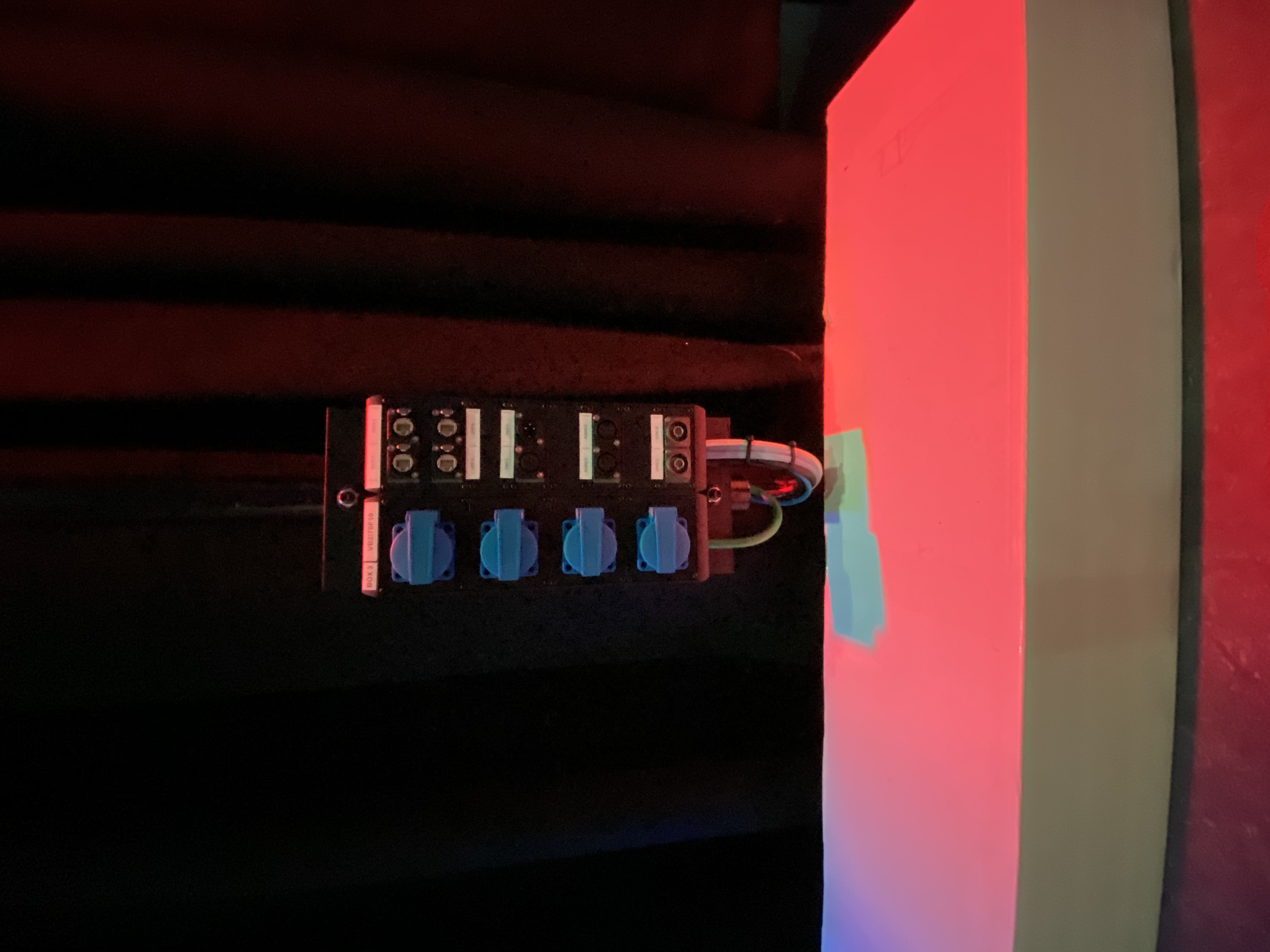} }}%
        \qquad
        \subfloat[\centering Patch Rack]{{\includegraphics[height=4cm]{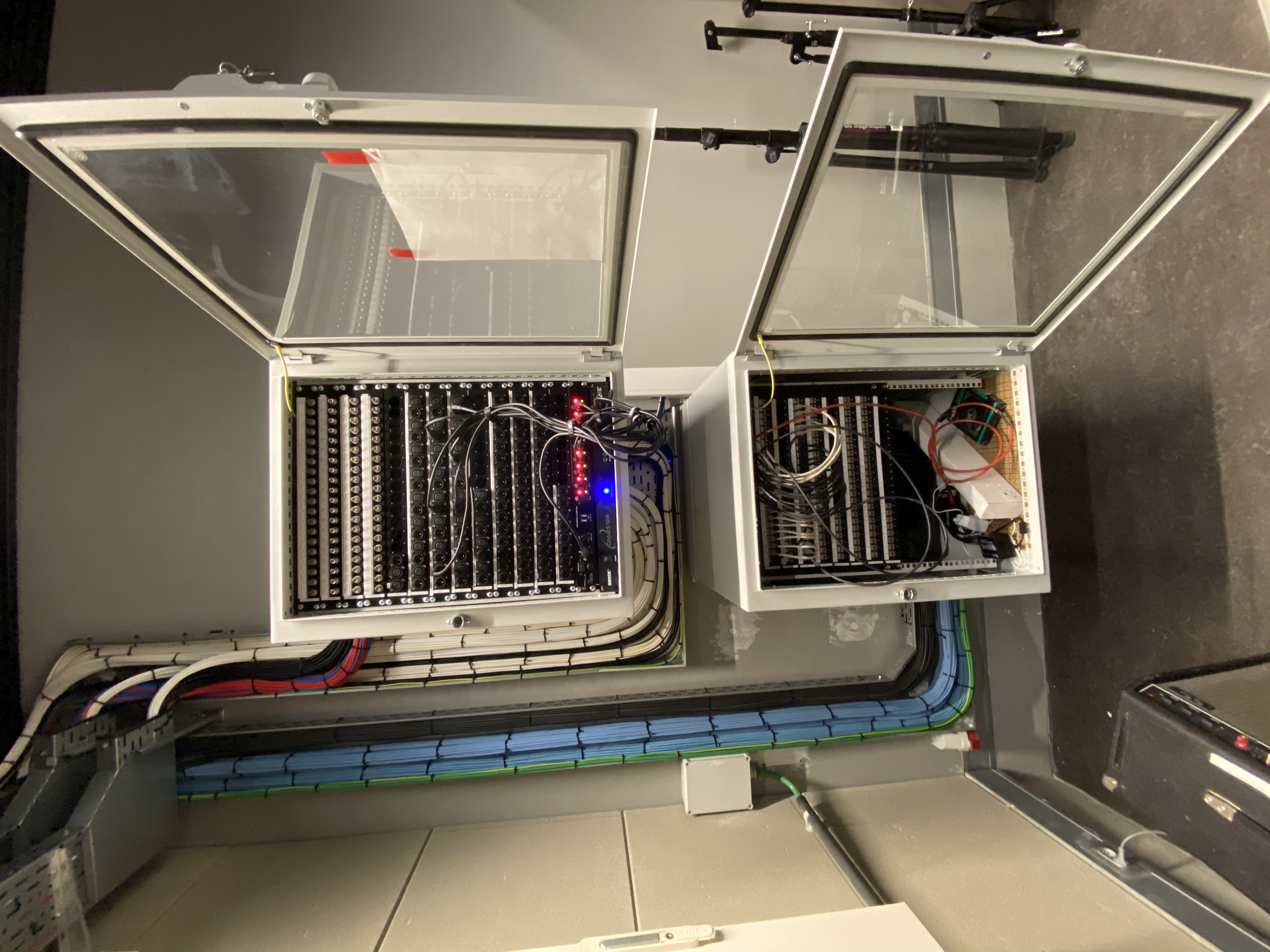} }}%
        \caption{Patching infrastructure}%
        \label{fig:patchbox}%
\end{figure}

In order to limit the length and amount of cables to be used in setting up new experiments, a large amount of patch boxes have been installed on both the fixed and moving trusses. All patch boxes deliver power, data, UTP, coaxial, DMX and XLR connections to several locations in the lab (Fig. \ref{fig:patchbox}a). Each of these patch-points are directly linked to the patch rack (Fig. \ref{fig:patchbox}b), from where connections towards the machine room or other patch points can be made. 

\section{Audiovisual equipment}

The Art and Science Interaction Lab is equipped with 80 individual speakers connected to a fully IP-based audio distribution system. The system is capable of delivering highly realistic spatial audio projection by making use of a dedicated audio wavefield processor. This audio system allows for accurate recreation and simulation of room acoustics. \\

In terms of visual modalities, the lab is equipped with 2 fully untethered HTC Vive Pro Eye, 2 tethered HTC Vive Pro and 2 Microsoft Hololens version 2 devices, allowing free roaming spanning the full 10m x 10m area and allowing for multi person AR/VR. A 7m x 4m acoustically transparent projection screen in combination with a 12.000 lumen 4K projector delivers compelling and high-end immersive visualizations. Both the audio and visual systems are connected to a powerful state-of-the-art processing backend.

\subsection{Audio equipment}

An immersive speaker system and a state-of-the-art sound processing backend allows compelling auditory experiences. Due to the acoustic insulation, sound reflections and deformations are vastly reduced in order to deliver the most accurate auditory stimuli to the listeners.

\subsubsection{Equipment list}

\begin{itemize}
    \item 80 calibrated speakers (8 subs, 2 speaker rings and an overhead speaker set)
    \item Barco IOSONO \cite{iosono} core wavefield synthesis system
    \item Fully IP-based Dante \cite{dante} audio infrastructure
    \item 10 amplifiers including built-in DSP for each individual speaker channel
    \item Highly flexible XLR and IP-audio patch possibilities
    \item Fully synced with 48kHz clock
    \item Software/frameworks: Ableton Live, Max/MSP, Ambisonics...
\end{itemize}

\subsubsection{Applications:}

\begin{figure}[!htbp]
    \centering
    \caption{Audio Applications - Musical Performance}
    \includegraphics[width=\linewidth]{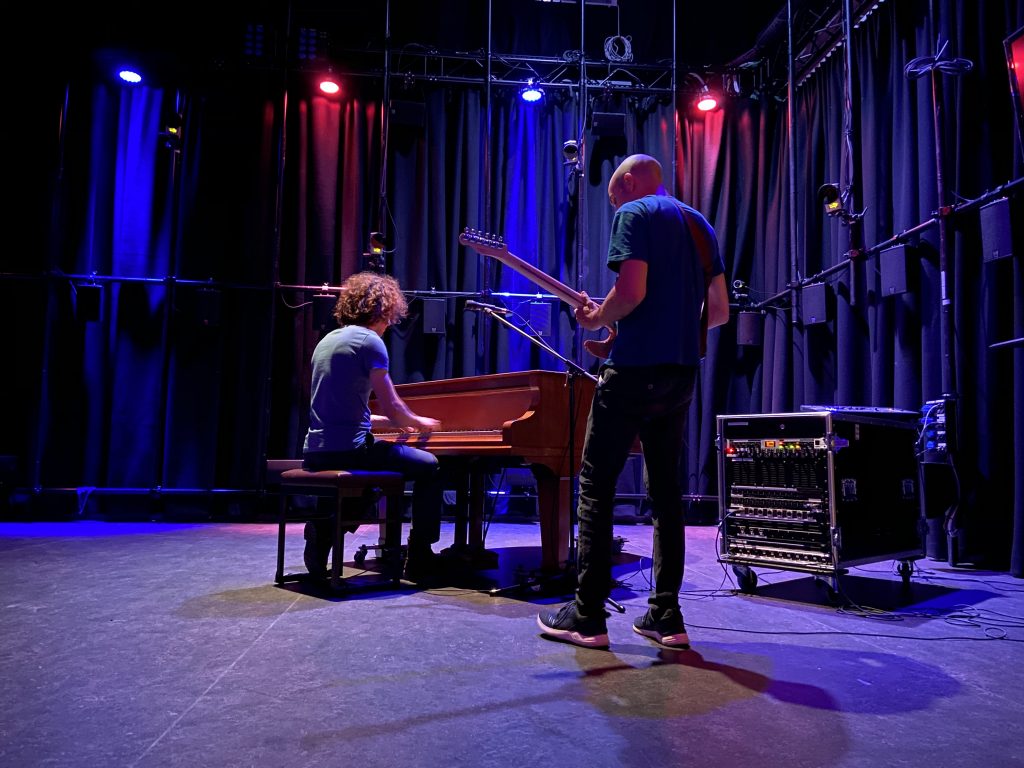}
    \label{fig:audioapplications}
\end{figure}

The audio installation in the Art and Science Interaction Lab allows for accurate object-based projection of 3D audio in space by using wavefield synthesis. This enables the (re-)creation of immersive auditory environments, bringing a myriad of application and research opportunities such as the recreation of multi-track musical performances and the use of interactive sound objects that can dynamically move in space. \\

Furthermore, such a system allows for the (re-)creation of real-life or simulated acoustic environments. This allows one to experience the acoustic properties of a certain location (e.g. concert hall, church, outdoors...) or the simulation of acoustic properties of future buildings or expositions.

\subsection{Video and Mixed AR/VR equipment}

The visual installation caters towards multi-user applications, with the focus on usability and freedom of movement. State-of-the-art virtual and augmented reality headsets are readily available in the lab. Furthermore, a high-resolution projection system is capable of delivering compelling grouped or single experiences.

\subsubsection{Equipment}

\begin{itemize}
    \item 2x HTC Vive Pro Eye untethered VR headsets with built-in eye tracking
    \item 2x HTC Vive Pro tethered VR headsets
    \item 2x Microsoft Hololens v2 AR glasses
    \item 7m x 4m acoustically transparent projection screen
    \item 12.000 lumens 4K projector with active stereo 3D and Extended Dynamic Range
    \item high-end rendering with latest NVIDIA GPU compute capabilities
\end{itemize}

\subsubsection{Applications}

\begin{figure}[!htbp]
    \centering
    \caption{Visual applications}
    \includegraphics[width=\linewidth]{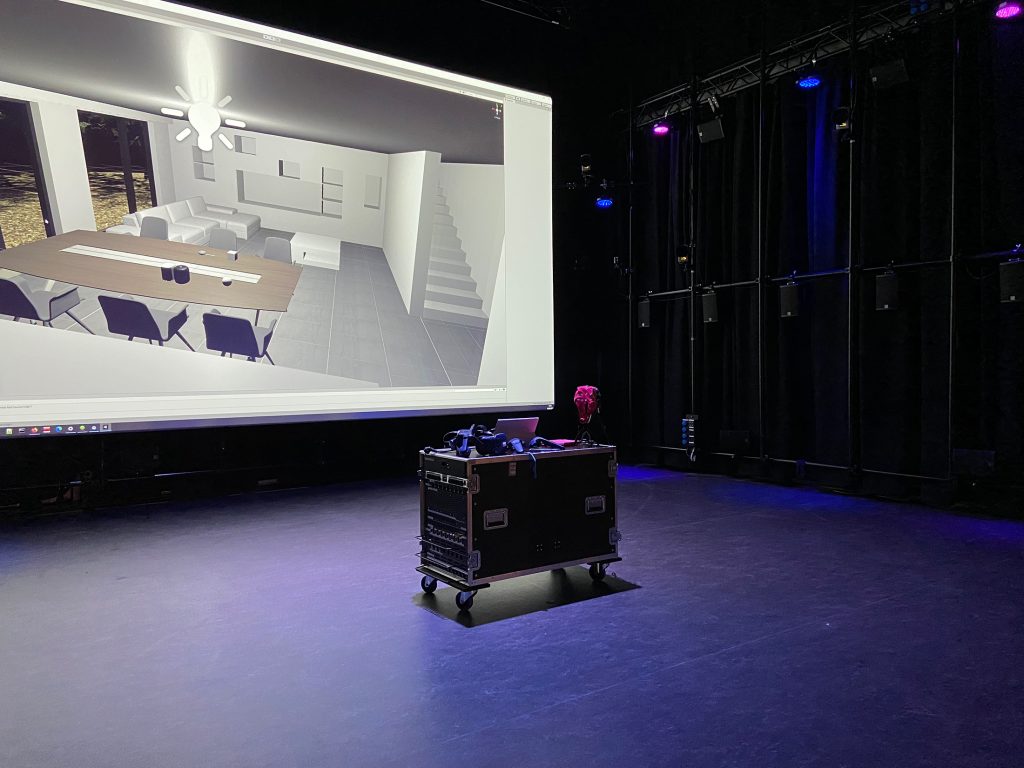}
    \label{fig:visualapplications}
\end{figure}

The visual systems integrated in the Art and Science Interaction Lab allow for the delivery of immersive and interactive virtual and augmented reality experiences while maintaining the highest degree of free movement possible. The state-of-the-art processing backend allows for real-time rendering and visualization of complex 6 degrees of freedom experiences (6DoF). \\

Secondly, the systems allow for multi-user VR and AR applications, allowing multiple users to interact naturally in a shared virtual environment. Integrated with the aforementioned positional audio and the motion capture system that will be described in the following section, these interactions can be of very high realism. \\

Lastly, the high resolution, high dynamic range projection allows for immersive context creation without the use of a virtual reality headset, as well as delivering compelling visual experiences to larger groups.

\subsection{Motion capture}

A dense Qualisys motion capture setup allows full body tracking of multiple users (up to 5) with an accuracy of $<$ 1mm\textsuperscript{3}, on a 81m\textsuperscript{2} floor area and a 5m configurable volume height. This allows detailed tracking of (multi) user movement. Furthermore, real-time integration with both the audio and visualization solutions allow for interactive audiovisual experiences based on human movement.

\subsubsection{Equipment}

\begin{itemize}
    \item Qualisys Oqus 7+ infrared cameras \cite{qualisysoqus} (14 fixed+ 4 mobile units)
    \item Qualisys Miqus Video cameras \cite{qualisysmiqus} (4 fixed + 1 mobile unit)
    \item Synced with the Art and Science Interaction Lab 120Hz clock signal
    \item High-end real-time processing backend allowing for real-time skeleton tracking and streaming
    \item Tracking compatible with VR headsets
\end{itemize}

\subsubsection{Applications}

\begin{figure}[!htbp]
    \centering
    \caption{Motion Capture - Qualisys Oqus 7+ IR camera}
    \includegraphics[width=\linewidth]{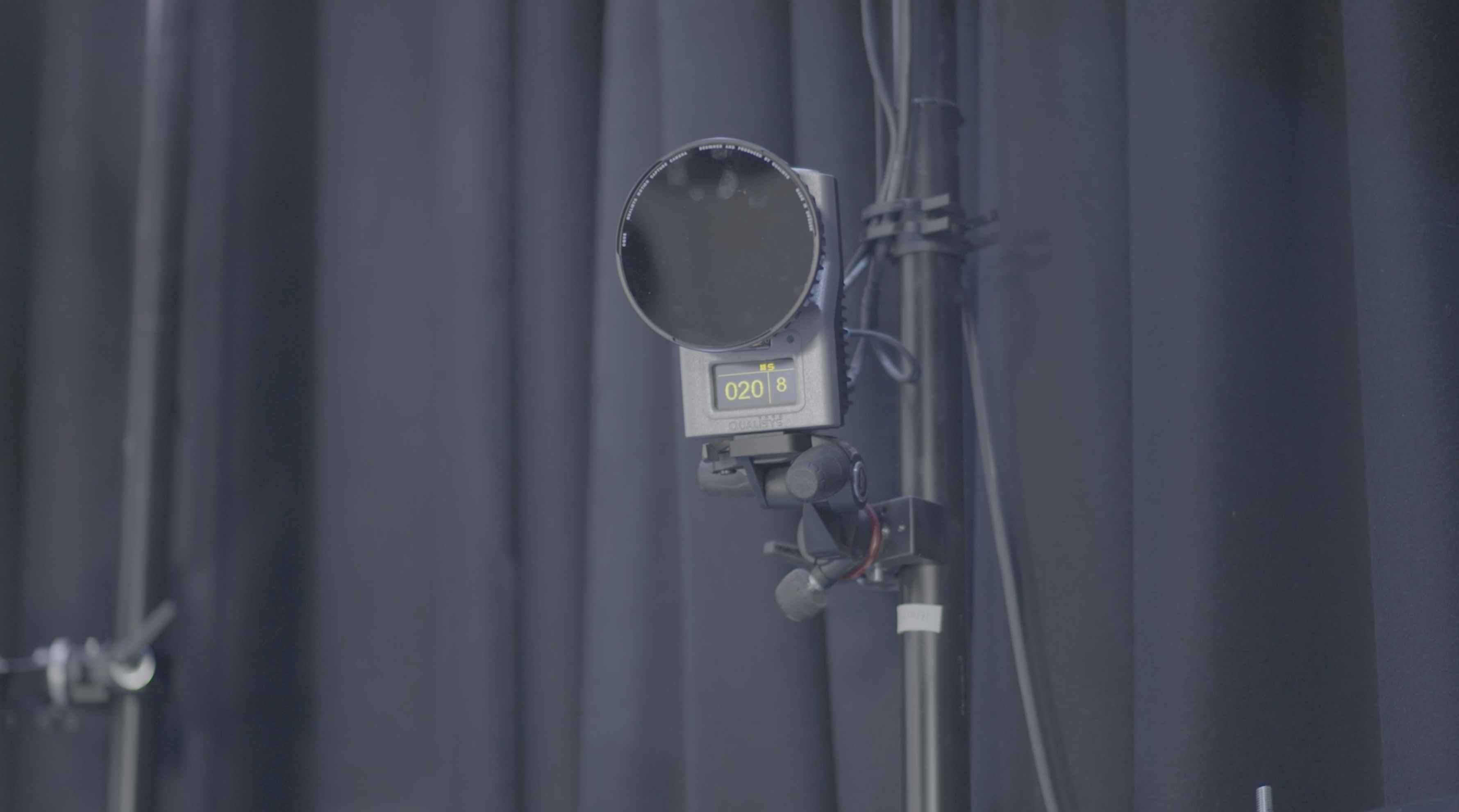}
    \label{fig:mocapapplications}
\end{figure}

The motion capture installation in the Art and Science Interaction Lab allows for highly accurate measurements of human behavior (motion) in interactive scenarios. The system allows for fine-granular motion tracking (e.g. finger tracking while playing an instrument, face expression tracking...) but also robust full body tracking of multiple users in the same area. \\

The system is capable of mapping marker data onto skeletons in order to animate avatars in VR and in real-time. This allows for accurate real-time representations of users in a virtual (e.g. VR) space. \\

Next to tracking users, the motion capture system allows for accurate 6 degrees-of-freedom tracking of objects. This by applying reflective markers to real-life objects. 

\section{Sensor equipment}

The Art and Science Interaction Lab is the go-to facility for interaction and user-experience research. The infrastructure provides a wide variety of synchronized sensors capable of measuring different aspects of an experience and is backed by a strong team of UX researchers. \\

The interdisciplinary team of the Art and Science Interaction Lab can deliver a detailed unraveling of user experiences and interactions, even those where users are not aware of. 

\subsubsection{Equipment:}

\begin{itemize}
    \item 2x clinical grade untethered EEG headsets (64 channels with active electrodes), which can be used in combination with our wireless VR headsets.
    \item Eye trackers, both built into VR headsets and standalone.
    \item Skin conductance sensors
    \item Heart rate sensors
    \item EMG sensors
    \item Synced with the Art and Science Interaction Lab 120Hz clock signal
\end{itemize}

\subsubsection{Applications:}

\begin{figure}[!htbp]
    \centering
    \caption{Sensor Applications - Physiological Analysis}
    \includegraphics[width=\linewidth]{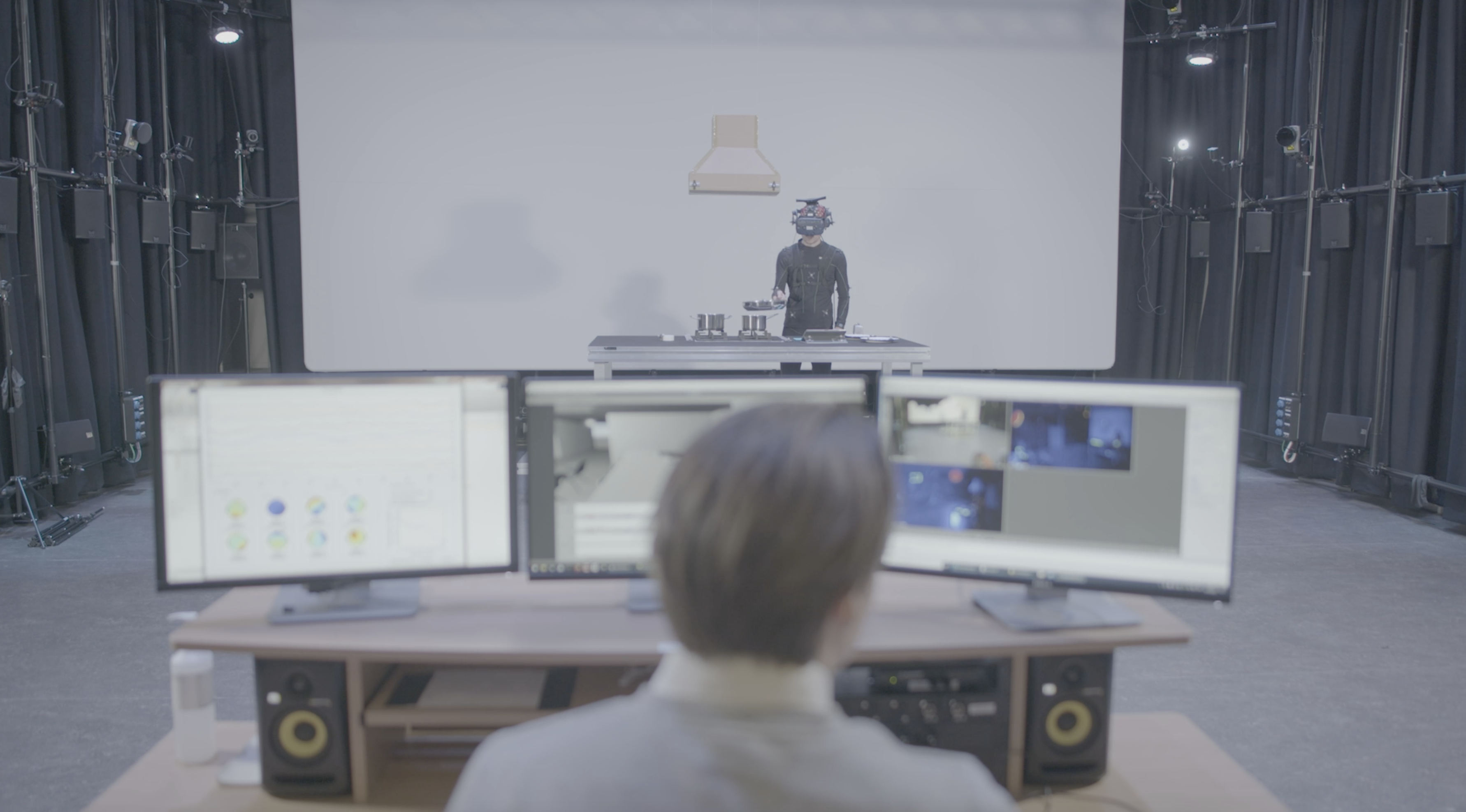}
    \label{fig:sensors}
\end{figure}

The available sensors in the Art and Science Interaction Lab and their integration and synchronization allow the measurement of a highly accurate physiological footprint of a user experience. By combining different sensors such as EEG, EMG, skin conductance... and the aforementioned motion capture system, physiological processes within human interactions and experiences can be unraveled on a granular scale. This analysis can be used to discover points of pain in a certain user-experience. \\

Furthermore, in a more artistic setting, real-time sensor data can be used in order to create novel audiovisual experiences based on real-time biofeedback.




\section{Interconnectivity}

The Art and Science Interaction Lab serves as a central hub towards different concert halls in the city center of Ghent. A 10Gbps dark fiber towards 'De Vooruit', 'De Minard' and the future 'Wintercircus' allows for real-time shared experiences between the Art and Science Interaction lab and three important cultural venues in Ghent. \\

Next to the cultural venues, the Art and Science Interaction Lab is also directly connected towards URGENT, the Ghent University Student radio, which also resides in De Krook. 

Lastly, the lab has a direct dark fiber link to an off-site back-up facility, delivering a secure and redundant data storage. 

\subsubsection{Applications:}

The direct link with the concert venues and URGENT radio allow for real-time streaming of multi-track audio between these facilities in order to enable the (re-)creation of a cultural experience in the Art and Science Interaction Lab as well as creating novel experiences based on real-time audio feeds. Furthermore, real-time motion capture allows for novel experiences where motion in the Art and Science Interaction Lab can be used in real-time in the cultural venues (e.g. the creation of virtual avatars or experiences).


%
\bibliographystyle{splncs04}
\bibliography{main}

\end{document}